\documentclass[12pt]{iopart}

\usepackage{graphicx}% Include figure files
\usepackage{braket}
\usepackage{dsfont}
\usepackage{amsfonts}
\usepackage{amssymb}
\usepackage{bm}

\usepackage{xcolor}
\usepackage{ulem}

\begin{document}

\title[Doubly-dressed states for near-field trapping and subwavelength lattice structuration]{Doubly-dressed states for near-field trapping and subwavelength lattice structuration}

\author{Maxime Bellouvet$^{1}$, Caroline Busquet$^{1,2}$, Jinyi Zhang$^{1}$, Philippe Lalanne$^{1}$, Philippe Bouyer$^{1}$ and Simon Bernon$^{1}$\footnote{http://www.coldatomsbordeaux.org/aufrons}}

\address{$^1$ LP2N, IOGS, CNRS and Universit\'e de Bordeaux, F-33400 Talence, France}
\address{$^2$ Muquans, Institut d'Optique, F-33400 Talence, France}
\ead{simon.bernon@institutoptique.fr}
\vspace{10pt}
\begin{indented}
\item[]January 2018
\end{indented}

\begin{abstract}
We propose a scheme to tailor nanostructured trapping potentials for ultracold atoms. Our trapping scheme combines an engineered extension of repulsive optical dipole forces at short distances and attractive Casimir-Polder forces at long distances between an atom and a nanostructured surface. This extended dipole force takes advantage of excited-state dressing by plasmonically-enhanced intensity to doubly dress the ground state and create a strongly repulsive potential with spatially tunable characteristics. In this work, we show that, under realistic experimental conditions, this method can be used to trap Rubidium atoms close to surfaces (tens of nanometers) or to realize nanostructured lattices with subwavelength periods. The influence of the various losses and heating rate mechanism in such traps is characterized. As an example we present a near-field optical lattice with 100nm period and study the tunability of lattice and trapping depths. Such lattices can enhance energy scales with interesting perspectives for the simulation of strongly-correlated physics. Our method can be extended to other atomic species and to other near-field hybrid systems where a strong atom-light interaction can be expected.
\end{abstract}

\section{Introduction}

Various hybrid quantum systems have been developed in the recent years with the idea that the overall combination of systems could exploit the properties of individual systems at best and overcome their individual performances \cite{Kurizki15}. In this spirit, an intensive research has been carried on various combinations such as for example NV centers or cold atom coupling to superconducting circuits processors for quantum memory applications \cite{Amsuss11,Kubo10,Diniz11,Bernon13}, magnetic sensing \cite{Weiss15,Bienfait16} or hybrid combinations involving macroscopic quantum systems such as mechanical oscillators \cite{Arcizet11,Kolkowitz12,Fang16,Vochezer18}. In this family of coupled quantum systems, ultracold atoms are special because of their very well-controlled intrinsic properties due to their strong decoupling from the environment and because they stand in gaseous form while other systems are predominantly solid-state devices. Transporting atoms close to a quantum counterpart is crucial for applications in quantum information and quantum simulation \cite{bloch12}. By increasing the coupling strength with, for instance, microwave superconducting photons \cite{Henschel17,Hattermann17}, optical photons \cite{Pappa15,Goban14,Sayrin15,Thompson13}, slow light \cite{Douglas15,Zang16} or plasmons \cite{stehle14}, we can devise efficient quantum memory or engineer surface-metiated atom-atom interactions \cite{gullans12,hood16,gonzalez15}. 
Tailoring  nanostructured optical lattice quantum simulators with tunable properties (energy scale, geometry, surface-mediated long-range interaction) will allow one to investigate new regimes in many-body physics. In this quest for exotic quantum phases (e.g., quantum antiferromagnetism \cite{RevModPhys.80.885}), the reduction of thermal entropy is a crucial challenge \cite{Bernier09,Hulet15,cheuk15,greif16,Mazurenko17}. The price to pay for such low temperature and entropy is a long thermalization time that will ultimately limit the experimental realization. Miniaturization of lattice spacing is a promising solution to speed up this dynamics. There is therefore a wide effort in the community to push the optical trapping technology to the nanometer scale \cite{stehle14,gullans12,gonzalez15,stehle11,Lukin13,chang14} or to generate subwavelength lattice spacing using multi-photon optical transitions \cite{PhysRevA.66.045402,PhysRevA.74.063622}, time-periodic modulation \cite{PhysRevLett.115.140401} or superconducting vortices \cite{PhysRevLett.111.145304}. Engineering cold atom hybrids offers promising perspectives but requires to interface quantum systems in different states of matter at very short distances, which still remains an experimental challenge.

In this paper, we present a novel trapping method that enables to trap and manipulate cold atoms below 100 nm from surfaces where the electromagnetic environment of cold atoms could be engineered. We show that our method can also be used to create subwavelength potentials with controllable parameters. Our method, that we refer as Doubly-Dressed State (DDS), is inspired by \cite{gonzalez15,chang14} where the authors take advantage of vacuum forces and material engineering to create near-field traps. Here, two off-resonant laser light fields interacting with the structured surface overcome the Casimir-Polder (CP) attraction at short distances \cite{casimir48}. The combination of the DDS repulsion and CP attraction forms a trap that follows the shape of the surface. A modulation of the CP potential can be generated with a periodically-structured surface, thus resulting in a lattice potential for cold atoms. In particular we study here the generation of a 1D lattice by periodically-spaced dielectric ridges in the spirit of \cite{gonzalez15}. Another scheme for the realization of surface traps with subwavelength optical lattices is published back-to-back \cite{Mildner18}.

In section \ref{sec:level1} we present the general arguments of the trapping method which combines optical and vacuum forces. In section \ref{sec:level2} we apply  the method to a stratified surface and characterize the trapping properties. We then extend it to a one-dimensional nanostructured surface where longitudinal and transverse trapping in a subwavelength lattice is demonstrated. In section \ref{sec:level3} we discuss the advantages and limitations of this method, to finally conclude in section \ref{sec:level4} with perspectives.

\section{\label{sec:level1}Trapping method}

%We consider here an atom in the vacuum close to a surface. 
In the absence of external field, an atom in vacuum close to a surface interacts with it because of quantum charge and field fluctuations, also known as the Casimir-Polder (CP) interaction \cite{casimir48}. It can be calculated by assimilating the atom to a fluctuating induced dipole with a moment $\mathbf{d}$, which creates an electromagnetic field that is reflected by the surface before interacting back with the dipole. This CP interaction is therefore characterized by an energy $U$ proportionnal to the atomic polarizability $\alpha$ and to the scattering Green tensor $\mathbf{G}^{(1)}(\mathbf{r},\mathbf{r},\omega)$. The latter describes the round-trip propagation (between the atom and the surface) of an electric field at frequency $\omega$ intially generated at the position of the atom $\mathbf{r}$. It contains all the optical properties of the surface. The CP force derived from the energy between two electric or paramagnetic objects is always attractive for an atom in the ground state.

\begin{figure}[h]
\centering
\includegraphics[scale=0.8]{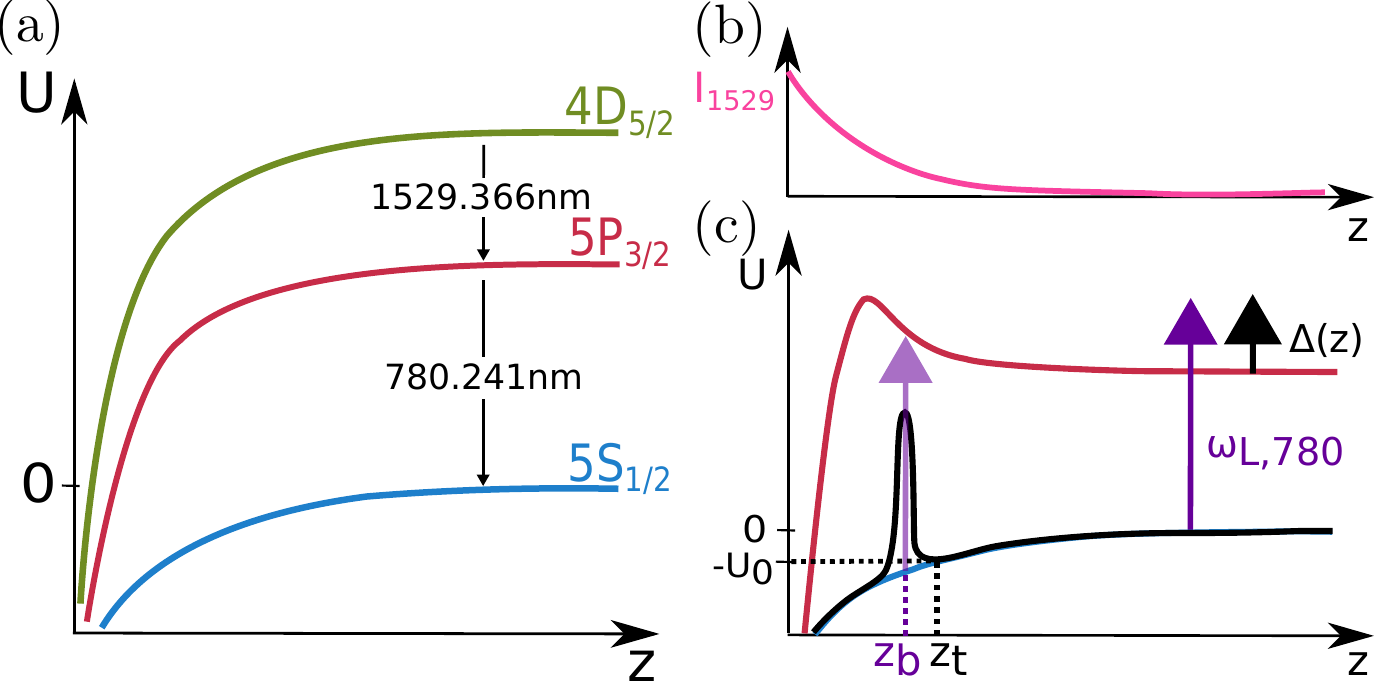}
\caption{\textbf{Trapping method.} \textbf{(a)} Quantum fluctuations between a $^{87}Rb$ atom and a surface (in $z<0$) give rise to an attractive CP potential. The force depends on the internal state of the atom. Here are represented the potentials for three different states: $5S_{1/2}$ (blue), $5P_{3/2}$ (red) and $4D_{5/2}$ (green). \textbf{(b)} We consider a spatially-varying (e.g., exponentially) light field with a maximum intensity at the surface-vacuum interface ($z=0$) and a frequency close to the transition $5P_{3/2}\rightarrow 4D_{5/2}$ (\textit{i.e.} $1529.366$~nm). \textbf{(c)} The modulation of the $5P$ state creates a spatially-varying frequency $\omega_{0}(z)$ for the  $5S_{1/2}\rightarrow 5P_{3/2}$ transition. By adding a second laser at a frequency $\omega_{L,780}$ (violet) detuned from this transition (\textit{i.e.} $\omega_{0,vac}=780.241$~nm in the vacuum) the atom is in a doubly-dressed state. The laser being resonant at $z_b$, the combination of CP attraction and DDS repulsion creates a trap at position $z_t$ with depth $U_0$.}
\label{fig1}
\end{figure}

We consider here a semi-infinite dielectric surface with a dielectric-vacuum interface at $z=0$. Fig.\ref{fig1}\textbf{(a)} sketches three internal states of a $^{87}Rb$ atom modulated by the vacuum forces only. In that case, the potentials are attractive and drag the atoms to the surface. We consider now a laser field with an intensity profile decaying from the surface as shown in Fig.\ref{fig1}\textbf{(b)}, which dresses the $5P_{3/2}$ state. This laser is tuned a few hundreds of GHz on the blue of the $5P_{3/2}$ to $4D_{5/2}$ resonance. The resulting AC Stark shift effect induces a strong repulsive potential for the $5P$ state while barely affecting the $5S_{1/2}$ state. The $5S$-$5P$ transition frequency $\omega_{0}(z)$ therefore depends on the distance $z$ from the surface. Double dressing is achieved by using a second laser at frequency $\omega_{L,780}$ tuned close to  the free space $5S-5P$ transition $\omega_{0,vac}$. The resulting AC Stark shift on the $5S$ state, is governed by the spatially-dependent detuning $\Delta(z)=\omega_{L,780}-\omega_{0}(z)$ (see Fig.\ref{fig1}\textbf{(c)}), which cancels at position $z_b$. This position directly depends on the laser frequency $\omega_{L,780}$. For positions $z>z_b$, the detuning $\Delta(z)$ is positive (blue) and the doubly-dressed potential expels the atoms away from the surface. As depicted in Fig.\ref{fig1}\textbf{(c)}, a trap is then formed by the contribution of the attractive CP and repulsive doubly-dressed state potentials.

The trapping potential is calculated by integrating the time derivative of the atomic momentum $\hat{p}_A$ over the trajectory of the atom \cite{Dalibard85} :
\begin{equation}
U(z)=\int_\infty^z dz' \left\langle\frac{d\hat{p}_A(z')}{dt}\right\rangle =-\int_\infty^z dz' \left[\rho_{ee}(z')\frac{dU_{5P}}{dz'}+\rho_{gg}(z')\frac{dU_{5S}}{dz'}\right]
\end{equation}
where $\rho_{ee}$ (resp. $\rho_{gg}$) is the population of the $5P$ (resp. $5S$) state obtained by solving the optical Bloch equations. This expression for the total potential states that both excited and ground state energy gradient contributions (state population $\times$ energy gradient) are opposite at the trap position $z_t$ : 
$\rho_{ee}(z_t)\left.\frac{dU_{5P}}{dz'}\right|_{z_t}=-\rho_{gg}(z_t)\left.\frac{dU_{5S}}{dz'}\right|_{z_t}$.

\section{\label{sec:level2}Results}

\subsection{Case of a planar and stratified surface}
\label{stratPart}

As shown later, a key ingredient of the DDS trapping method is the amplitude of the excited-state energy shift. As explained in section \ref{sec:level1}, the decaying profile of the 1529nm-laser intensity directly imprints the energy shift of the 5P state. The spatial variation of intensity (gradient) must be as strong as possible to create a sufficient barrier to trap the atoms. To create an enhanced exponentially-decaying field in the vacuum, we take advantage of Surface Plasmon Resonance (SPR)  that occurs at a metal-dielectric interface of a stratified surface when the angle of incidence is $\theta_{SPR}=\arcsin \left(n_{eff}/n_i\right)$ where $n_i$ is the refractive index in the incident medium and $n_{eff}$ is the effective index of the Surface Plasmon Polariton (SPP) mode. 

\begin{figure}[h]
\centering
\includegraphics[scale=1]{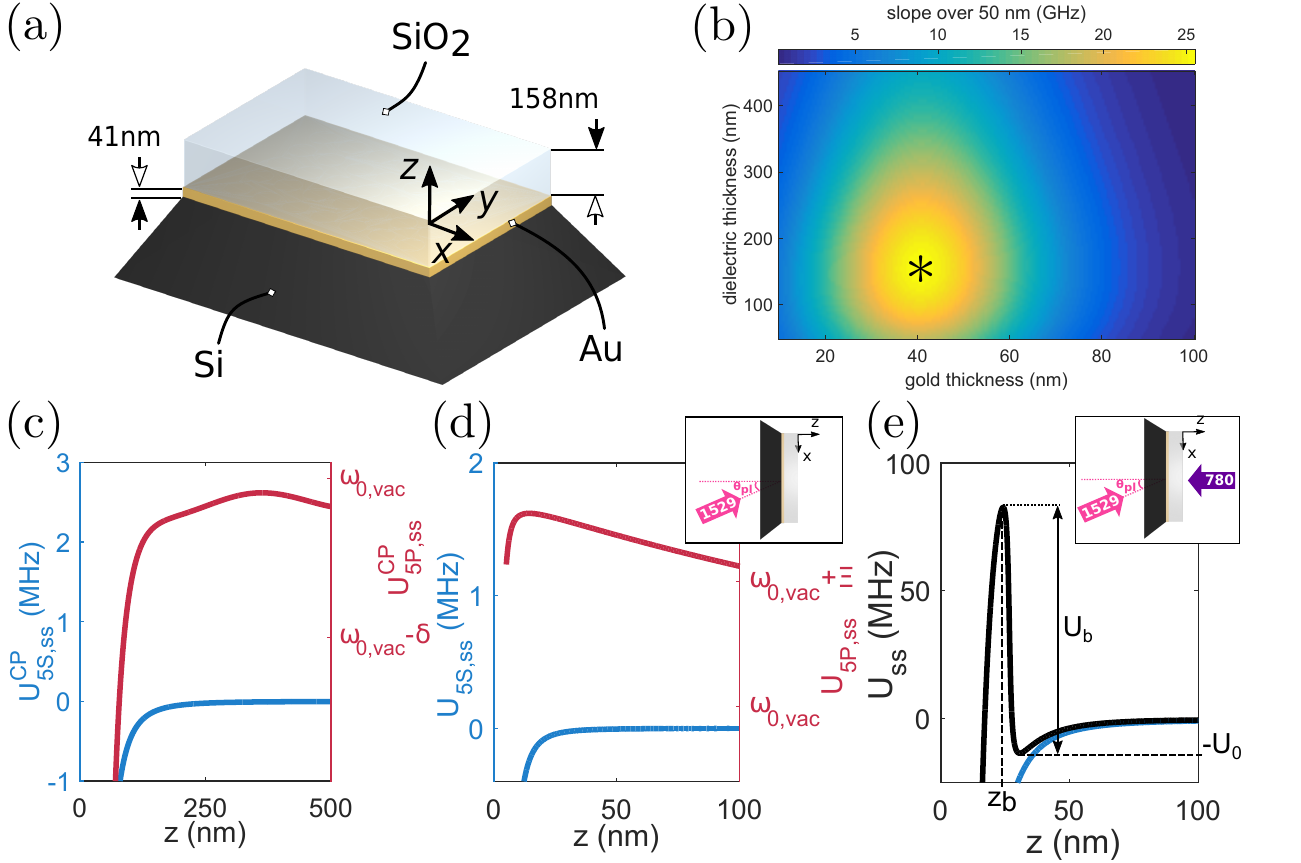}
\caption{\textbf{Trapping atoms in the near field.} \textbf{(a)} Representation of the stratified surface composed of a 158nm $SiO_2$ layer and a 41nm $Au$ layer deposited on a $Si$ substrate.  \textbf{(b)} The dimensions have been optimized by maximizing the gradient of the 1529nm-laser intensity over 50 nm (from $z$=50 to 100nm). The asterisk symbol correponds to the final configuration. \textbf{(c)} CP potentials for the $5S$ (blue) and $5P$ (red) states for a $^{87}Rb$ atom at a position $z$ from the stratified medium. $U_{5P,sm}^{CP}(z)$ is expressed in $\omega_{0,vac}$ which corresponds to the frequency of the $5S\rightarrow5P$ transition in vacuum (\textit{i.e.} $\omega_{0,vac}/2\pi \simeq 384.230$~THz). Here $\delta=5$~MHz. \textbf{(d)} Potentials (optical+CP) for the $5S$ (blue) and $5P$ (red) states when 1529nm laser (pink, see inset) is shined on the surface, with $I_{1529}=6.4\mu W/\mu m^2$. Here $\Xi=20$~GHz. \textbf{(e)} When the 780nm laser is added (violet in the inset) a barrier of potential with a height $U_b$ is created at $z_b$ in the total potential (solid black). The depth of the trap is defined by $U_0$. The parameters for the 780nm laser are $\Omega_R/2\pi\sim 132$MHz and $\Delta_0/2\pi\simeq 30$GHz.}
\label{fig2}
\end{figure}

The surface geometry studied in this section is given in Fig.\ref{fig2}\textbf{(a)}. The geometrical parameters of the stratified surface are chosen by maximizing the 1529nm-intensity decay  from 50~nm to 100~nm as shown in Fig.\ref{fig2}\textbf{(b)}. Given the optimization, we take a stratified structure composed of a dielectric layer ($SiO_2$) with a thickness of 158~nm and a 41nm layer of gold deposited on a $Si$ substrate. The lasers intensities are determined using transfer matrix method \cite{yeh98} implemented with Reticolo software \cite{reticolo}. 

The CP potentials are calculated using the multipolar coupling approach \cite{Buhmann04}. The thermal effects are neglected considering the distances at which the atoms are trapped ($z_A << \lambda_T=\hbar c/k_BT \simeq 7.6\mu m $, $z_A$ being the atom position and $\lambda_T$ the thermal wavelength). For an atom in a state $n$ ($n\equiv 5S_{1/2}$ or $n\equiv 5P_{3/2}$), the interaction with the stratified surface is described by a potential that can be split into two terms :
\begin{equation}
U_{n,ss}^{CP}(z_A)=U_{n,ss}^{nres}(z_A)+U_{n,ss}^{res}(z_A),
\label{atom_stratifiedSurface_interaction}
\end{equation}
where
\begin{equation}
U_{n,ss}^{nres}(z_A)=\frac{\hbar\mu_0}{2\pi}\int_0^\infty d\xi \xi^2 \Tr\left[\boldsymbol{\alpha}^{(n)}(i\xi)\cdot \mathbf{G}^{(1)}(z_{A},z_A,i\xi)\right]
\label{CPPotentialsNResPart}
\end{equation}
is the non-resonant term. The second term, named resonant term, is non-zero if the atom is excited ($5P$) and is given by
\begin{equation}
U_{n,ss}^{res}(z_A)=-\mu_0\sum_{n<m}\omega_{nk}^2 \mathbf{d}_{mn} \cdot \mathrm{Re}\mathbf{G}^{(1)}(z_{A},z_{A},\omega)\cdot\mathbf{d}_{nm}.
\label{CPPotentialsResPart}
\end{equation}
By considering a $^{87}Rb$ atom with an isotropic polarizability $\boldsymbol{\alpha}^{(n)}(i\xi)=\alpha^{(n)}(i\xi)\mathds{1}$, these two expressions can be written as
\begin{equation}
\fl
U_{n,ss}^{nres}(z_A)=\frac{\hbar\mu_0}{8\pi^2}\int_0^\infty \xi^2d\xi\alpha_{Rb}^{(n)}(i\xi)\int_0^\infty dk^{\parallel}\frac{k^{\parallel}}{\kappa^\perp}e^{-2\kappa^\perp z_A}\left\{r_{s}(k^{\parallel},i\xi)+\left(1-\frac{2\kappa^\perp{}^2c^2}{\xi^2}\right)r_{p}(k^{\parallel},i\xi)\right\}
\label{atom_stratifiedSurface_interaction_nresTermV2}
\end{equation}
and
\begin{eqnarray}
\fl
U_{n,ss}^{res}(z_A)=&\frac{\mu_0}{12\pi}\omega_{0,vac}^{2}|\mathbf{d}_{0}|^2\int_0^{\omega_{0}/c} dk^\perp\left\{\mathrm{Im}\left(e^{2ik^\perp z_A}r_{s}\right)+\left(1-\frac{2k^\perp{}^2c^2}{\omega_{0,vac}^{2}}\right)\mathrm{Im}\left(e^{2ik^\perp z_A}r_{p}\right)\right\}\nonumber\\
&-\frac{\mu_0}{12\pi}\omega_{0,vac}^{2}|\mathbf{d}_{0}|^2\int_0^{\infty} d\kappa^\perp e^{-2\kappa^\perp z_A}\left\{\mathrm{Re}\left(r_{s}\right)+\left(1+\frac{2\kappa^\perp{}^2c^2}{\omega_{0,vac}^{2}}\right)\mathrm{Re}\left(r_{p}\right)\right\}.
\label{atom_stratifiedSurface_interaction_resTermV2}
\end{eqnarray}
For numerical reasons, the integrations are performed over imaginary frequencies $\xi$ ($\omega=i\xi$) for the non-resonant term  (\ref{atom_stratifiedSurface_interaction_nresTermV2}) and over imaginary wavevectors $\kappa$ ($k=i\kappa$) for the resonant term (\ref{atom_stratifiedSurface_interaction_resTermV2}).
We define longitudinal and transverse parts for the wavevector as follows: $\mathbf{k}= \mathbf{k}^\parallel+\mathbf{k}^\perp$ where $\mathbf{k}^\parallel=k^\parallel\mathbf{\hat{z}}$ and $\mathbf{k}^\perp=k_x\mathbf{\hat{x}}+k_y\mathbf{\hat{y}}$. $r_{s}$ (resp. $r_{p}$) is the amplitude reflection coefficient of the stratified surface for $s$ (resp. $p$) polarization defined by the electric (resp. magnetic) field being transverse to the plane of incidence. $|\mathbf{d}_{0}|$ and $\omega_{0,vac}$ are respectively the electric dipole moment and the frequency of the $5S-5P$ transition ($|\mathbf{d}_{0}|=5.977ea_0$ and $\lambda_{0}=2\pi c/\omega_{0}^{vac}=780.241$~nm for $^{87}Rb$). In this planar and stratified example, the coefficients $r_s$ and $r_p$ are calculated with Fresnel equations. The dynamic polarizability for an isotropic atom in a state $n$ at imaginary frequencies is given by \cite{Arora07}
\begin{equation}
%\fl
\alpha^{(n)}(i\xi;J_n)=\frac{2}{3\hbar \left(J_n+1\right)}\sum_{J_m} \frac{\omega_{mn}\left|\langle J_n ||\mathbf{d}||J_m\rangle\right|^2}{\left(\omega_{mn}^2+\xi^2\right)},
\label{polarizability}
\end{equation}
where the sum is carried over all possible $n\rightarrow m$ transitions with a frequency $\omega_{mn}$ and a reduced dipole moment $\left|\langle J_n ||\mathbf{d}||J_m\rangle\right|$. $J$ denotes the total electronic angular momentum of the state. The calculated CP potentials are plotted in Fig.\ref{fig2}\textbf{(c)} and present an oscillation for the excited-state modulation due to the possible spontaneously-emitted real photon. Both potentials are attractive for atoms at position $z_A< 300$~nm from the surface.

\begin{figure}[h]
\centering
\includegraphics[scale=0.5]{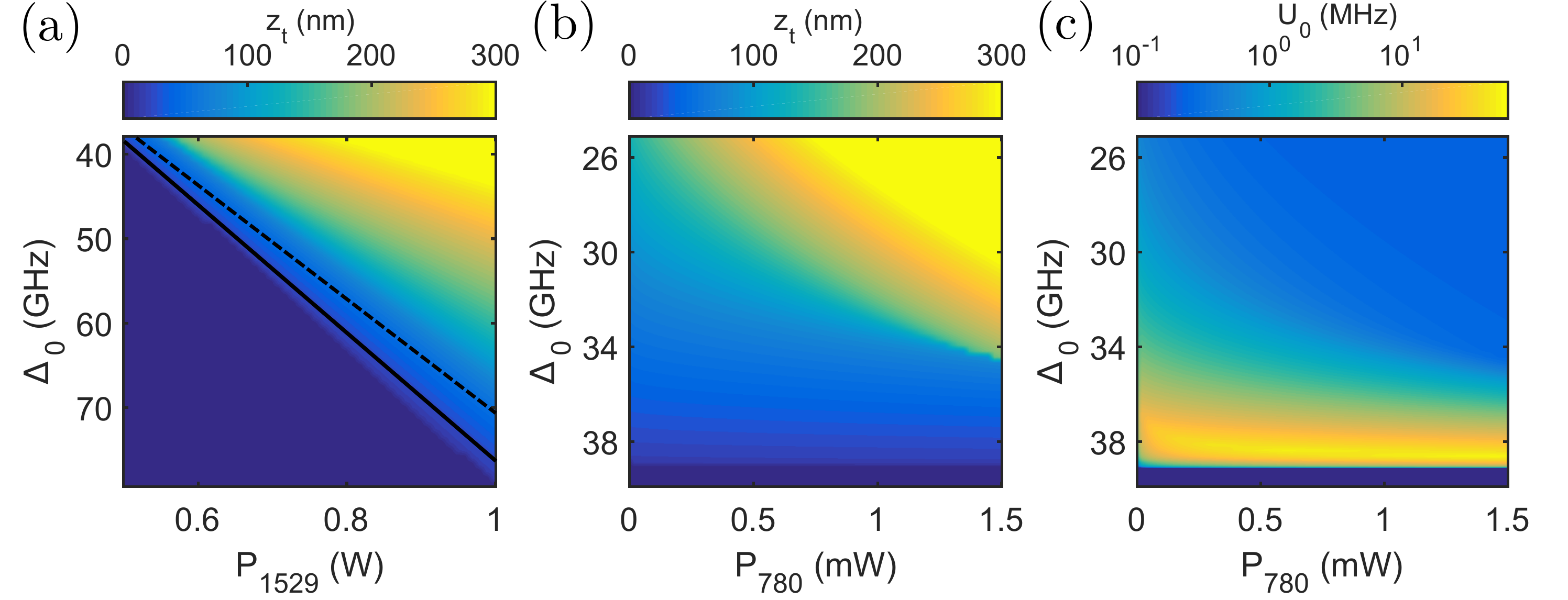}
\caption{\textbf{Tuning the geometrical parameters.} Position of the trap $z_t$ by scanning the 780nm-laser frequency and the powers of the \textbf{(a)} 1529nm laser and \textbf{(b)} 780nm laser. Both lasers have a waist of $200\mu m$. \textbf{(c)} Depth of the trap $U_0$ \textit{versus} the 780nm-laser beam free parameters : detuning $\Delta_0 = \omega_{L,780}-\omega_{0,vac}$ (vertical) and power $P_{780}$ (horizontal). The full and dashed lines in \textbf{(a)} correspond to constant trapping positions $z_t=25$ and 50~nm, respectively. The trap characteristics for these two positions are depicted in Fig.\ref{fig4}.}
\label{fig3}
\end{figure}

We now illuminate the rear side of the surface at the plasmonic angle $\theta_{SPR}\sim 18.6^\circ$ with laser light at a wavelength $\lambda_{L,1529}=1529.34$~nm, blue-detuned by 26 pm from the $5P-4D$ transition. As shown in Fig.\ref{fig2}\textbf{(d)}, using an incident power of 400 mW and a waist of 200 $\mu$m, we obtain a maximum SPP intensity at the dielectric-vacuum interface ($z=0$) of $610~\mu W/\mu m^2$ (100 SPP enhancement factor) corresponding to a maximal AC Stark shift of the $5P$ state of $\sim$ 31 GHz. This AC Stark shift is calculated using strong-field approximation (fine structure basis), justified by the SPP enhancement. The laser at $\lambda_{L,780}=780.201$~nm, \textit{i.e.} $\Delta_0=\omega_{L,780}-\omega_{0,vac}\simeq 2\pi\cdot30$~GHz, with a power of 200~mW in a 200~$\mu m$ waist is retroreflected on the surface (see inset in Fig.\ref{fig2}\textbf{(e)}). The detuning $\Delta(z)$ is zero at distance $z_b=24$~nm (resonant position). In this simulation we use a spatially homogeneous Rabi frequency $\Omega_R /2\pi \sim 132$~MHz and we obtain the total potential represented in Fig.\ref{fig2}\textbf{(e)}, with a trapping depth $U_0\simeq 13.5$~MHz and a trapping position $z_t\simeq 31$~nm. The geometrical properties of the trap such as the position and the depth can be adjusted with the powers of both lasers (1529 and 780 nm) and the frequency of the 780nm laser (Fig.\ref{fig3}).

As in \cite{chang14}, we evaluate the performance of our configuration by considering the four following characteristic times~: the exit time $\tau_{out}$, the tunneling time $\tau_t$, the anti-damping time $\tau_{ad}$ and the trap oscillation time $\tau_a$. We neglect the effect of transient heating \cite{chang14}. The exit time is due to scattering heating. It is defined as $\tau_{out}=\frac{dE}{dt}|E_g|^{-1}$ where $dE/dt = \hbar^2k_{eff}^2/(2m)\Gamma_{sc}$ is the energy change for a duration $dt$ due to the absorption by the atom with a mass $m$ of a photon with an effective momentum $\hbar k_{eff}$, at a scattering rate $\Gamma_{sc}=\Gamma_0\rho_{ee}=\tau_{sc}^{-1}$. $E_g$ is the ground-state-binding energy determined along with the wavefunction by using imaginary time propagation method. The tunneling time is due to the possibility for the atom to escape the trap by tunneling towards the surface. This characteristic time is calculated within the WKB approximation \cite{griffiths}. The anti-damping time corresponds to the time for the energy increase due to blue-transition heating to equal the energy $E_g$. In our case it is defined by $\tau_{ad}=\log\left[\left(|E_g|+\Delta p^2/(2m)\right)/\left(\Delta p^2/(2m)\right)\right]/2\beta$ where $\beta=-4\omega_r \frac{\Delta(z_t)\Gamma_{sc}}{\hbar k_0^2 \left|\Delta_c(z_t)\right|^4}\left.\frac{dU_{5P}}{dz}\right|_{z_t}\left.\frac{d\Delta}{dz}\right|_{z_t}$ is the anti-damping rate with $\Delta_c(z)=\Delta(z)+i\Gamma_0/2$ and the recoil energy $\omega_r$ \cite{chang14}. We note $\Delta z$ and $\Delta p$ the position and momentum standard deviations of the trapping ground state respectively. The trap oscillation time $\tau_a = m\Delta z/\Delta p$ represents a characteristic adiabatic time.

\begin{figure}[h]
\centering
\includegraphics[scale=0.35]{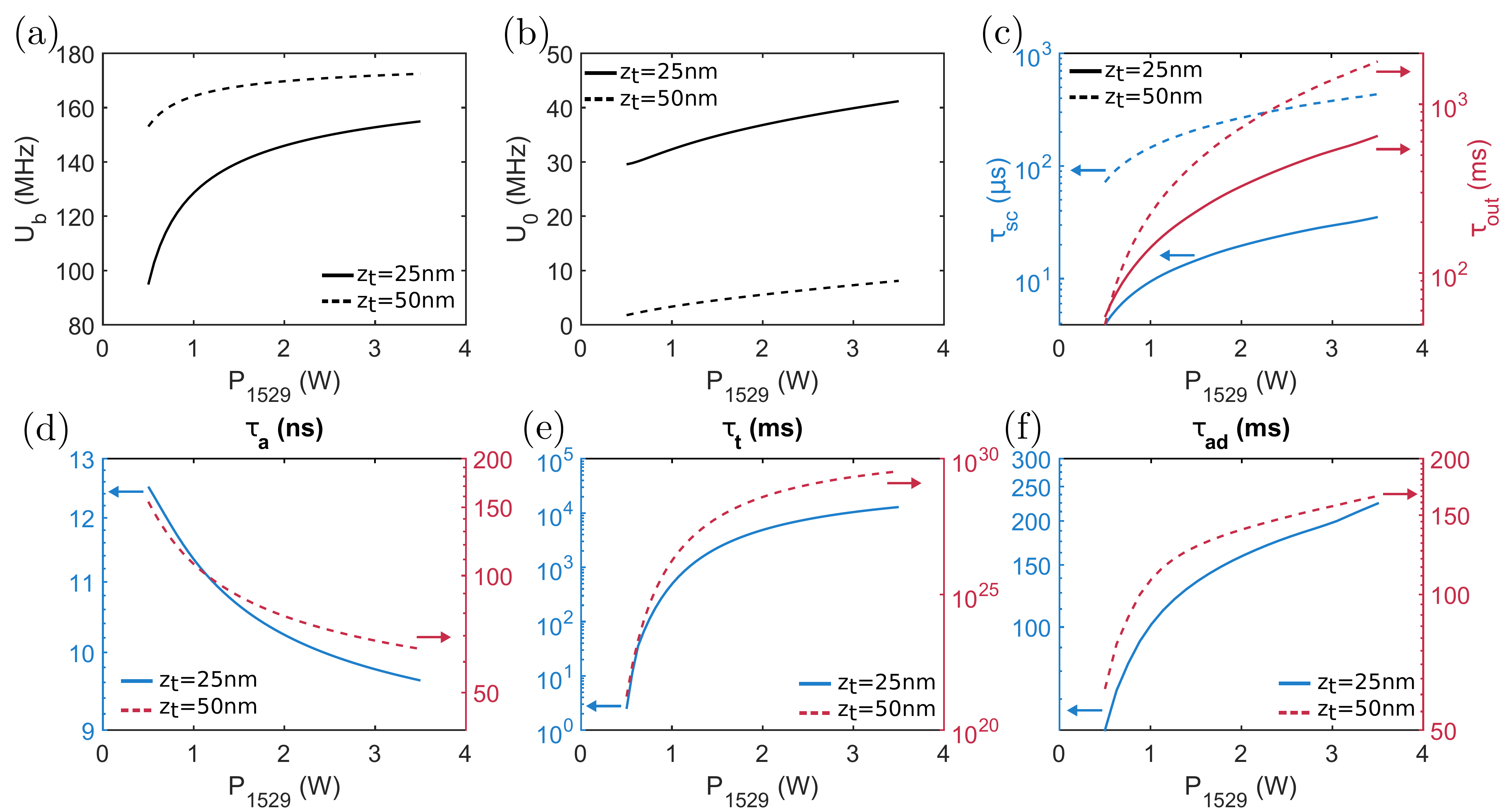}
\caption{\textbf{Trap characteristics.} The tuning of the 1529nm-laser power allows one to control the geometrical properties of the trap such as \textbf{(a)} the height of the barrier $U_b$ and \textbf{(b)} the depth of the trap $U_0$, or the different characteristic times such as \textbf{(c)} the scattering rate $\tau_{sc}$ and the exit time $\tau_{out}$ due to scattering heating, \textbf{(d)} the oscillation time $\tau_a$ (adiabaticity), \textbf{(e)} the tunneling time $\tau_t$ and \textbf{(f)} the anti-damping time $\tau_{ad}$. From \textbf{(a)} to \textbf{(f)} : the trap positions $z_t=25$nm and $z_t=50$nm are represented by the full and dashed lines, respectively. Here we took $\Omega_R/2\pi\sim 162$MHz. }
\label{fig4}
\end{figure}
In Fig.\ref{fig4}, we plot the trapping depth and barrier height as well as the characteristic evolution timescales for two trapping positions (corresponding to the dashed and full lines in Fig.\ref{fig3}\textbf{(a)}) as a function of the 1529nm-laser power. The trapping depth, which is mostly controlled by the ground-state ($5S$) CP potential, strongly increases while reducing the trap-to-surface distance (Fig.\ref{fig4}\textbf{(b)}). As shown in Fig.\ref{fig4}\textbf{(c)},\textbf{(e)},\textbf{(f)}, all contributions to the trap lifetime $\tau=\left(\tau_{out}^{-1}+\tau_{ad}^{-1}+\tau_{t}^{-1}\right)^{-1}$ increase together with the 1529nm-laser radiation. The 1529nm power ($P_{1529}$) is therefore a key parameter that tunes the energy gradient of the $5P$ state and optimizes the trap characteristics. Experimentally, the optimal power (maximal) will be given by the heat capacity and absorption of the surface at 1529nm. In the range of experimental parameters studied here and for atom-to-surface distances around $z_t = 50$nm, the exit and anti-damping times are the limiting trapping times, which increase together with $P_{1529}$. At higher intensities and for both trapping positions presented here, anti-damping heating becomes the limiting lifetime effect.

Within the WKB approximation, the tunneling time is exponentially dependent on the barrier height and width. For similar barrier heights (Fig.\ref{fig4}\textbf{(a)}), the reduction of barrier width  at shorter distances has a strong impact on the tunneling time (20-orders-of-magnitude decrease) as we can see in Fig.\ref{fig4}\textbf{(e)}. At short atom-to-surface distances, tunneling through the barrier will therefore become the limiting lifetime effect.

\subsection{Case of a 1D grating}

\begin{figure}[h]
\centering
\includegraphics[scale=0.4]{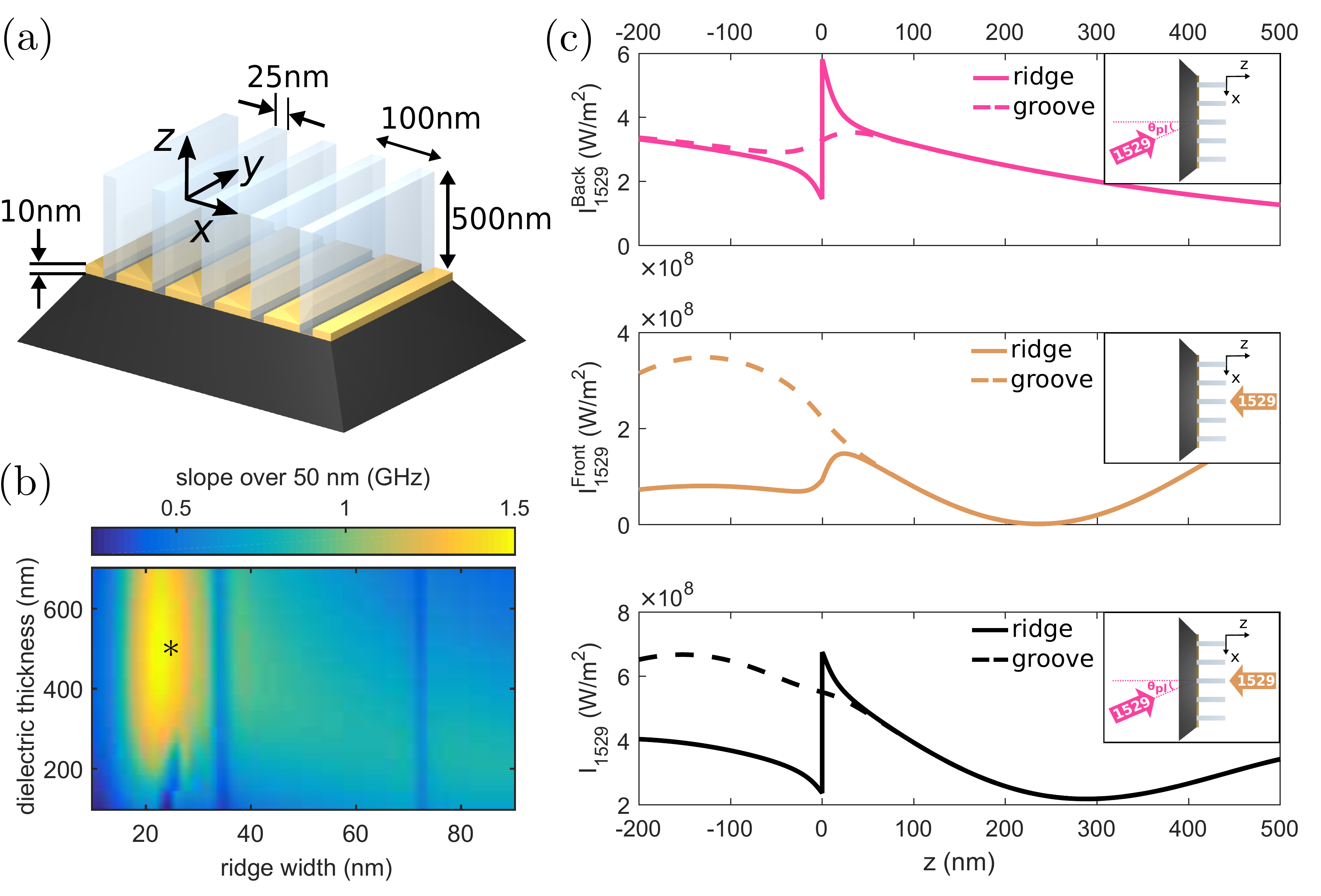}
\caption{\textbf{1D lattice structuration.} \textbf{(a)} The stratified surface is modulated by incorporating $SiO_2$ ridges. \textbf{(b)} The geometry is optimized by maximizing the intensity gradient over $50$nm. The asterisk symbol correponds to the final configuration : an optimal $SiO2$ thickness of 500nm, a ridge width of 25nm and a $Au$ film of 10nm. \textbf{(c)} The presence of a bump very close to the surface in the 1529-nm intensity distribution above a groove (dashed pink line) unlike above a ridge (full pink line) when sending the 1529nm laser backwards (in pink in the inset) is canceled out by adding a second beam at the same frequency but in the opposite direction (orange, see inset) to finally obtain the spatially-decaying intensity profiles at any position $x$ (black). For the three panels in \textbf{(c)} we took $P_{1529}^{Back}=1$W and $\alpha_{1529}=5$. Electromagnetic fields, which are p-polarized, are calculated with Reticolo software \cite{reticolo}.}
\label{fig5}
\end{figure}

In this section, we extend the previously described method to a nanostructured surface which modulates the CP potentials in the transverse direction (\textit{i.e.} ($xOy$) plane), therefore tailoring the trapping potential. Hereafter, we restrict our example to a 1D periodic nanostructuration, sketched in Fig.\ref{fig5}\textbf{(a)}, to create  a lattice potential with spacing fixed at $\ell=100$~nm. Using the same optimization procedure as in the case of the planar stratified structure, we optimize the geometry by maximizing the intensity decay of the 1529nm laser in front of a ridge (Fig.\ref{fig5}\textbf{(b)}). This led to a dielectric thickness of 500~nm, a ridge width of 25~nm and a metal thickness of 10~nm (optimization not shown).

The CP potentials for a structured surface are calculated using equations (\ref{CPPotentialsNResPart}) and (\ref{CPPotentialsResPart}) in which the scattering Green tensor for a one-dimensional grating is given by
\begin{equation}
\fl
\mathbf{G}^{(1)}(\mathbf{r}_A,\mathbf{r}_A,\omega)=\frac{i}{8\pi^2}\int_{-\pi/\ell}^{\pi/\ell} dk_x \sum_{m,n}e^{i\frac{2\pi}{\ell}(m-n)x_A}\int_{-\infty}^{\infty} dk_y \sum_{\sigma,\sigma'=s,p} \frac{e^{i(k_z^m+k_z^n)z_A}}{k_z^n}\mathbf{e}_{m+}^{\sigma}r_{mn}^{\sigma \sigma'}\mathbf{e}_{n-}^{\sigma'}
\end{equation}
where $\mathbf{r}_A=(x_A,0,z_A)$ is the atom position. $\mathbf{e}_{n-}^{\sigma'}$ is the unit vector describing the field that propagates along $-z$ in the $n^{th}$ order with a polarization $\sigma'$, and scatters out of the surface in the $m^{th}$ order with polarization $\sigma$, $r_{mn}^{\sigma \sigma'}$ being the reflection coefficient. The reflected field is then described by $\mathbf{e}_{m+}^{\sigma}$ and propagates along $+z$. The integration over $k_x$ is performed in the first Brillouin zone $[-\pi/\ell,+\pi/\ell]$ where $\ell$ is defined as the period of the grating. The amplitude reflection coefficients $r_{mn}^{\sigma \sigma'}$ at imaginary frequencies for the ground state CP potential are calculated by implementing Fourier modal method \cite{guerout13,contrerasreyes10,Buhmann:2015rka} and the ones at real frequencies for the excited state CP potential are calculated using Reticolo software.

The optical scheme uses here two laser beams at $1529$~nm : one being shined from the back at frequency $\omega_{1529}^{Back}$ that excites SPPs (noted '$Back$') and one shined from the front at frequency $\omega_{1529}^{Back}+\delta$ (noted '$Front$'). As shown in Fig.\ref{fig5}\textbf{(c)} top, the \textit{Back} laser does not create a decaying intensity profile in front of a groove.  This is detrimental to trap atoms at this position. The \textit{Front} laser produces the inverted profiles (Fig.\ref{fig5}\textbf{(c)} middle). By having a slight frequency mismatch $\delta$ the interferences can be time-averaged and the total contribution to the potential is given by the sum of both intensities. As shown in Fig.\ref{fig5}\textbf{(c)} bottom, the total intensity has a rapidly-decaying profile at any transverse position. We define $\alpha_{1529}=P_{152 9}^{Front}/P_{1529}^{Back}$ as the ratio of power between the \textit{Front} and \textit{Back} lasers.

\begin{figure}[h]
\centering
\includegraphics[scale=0.5]{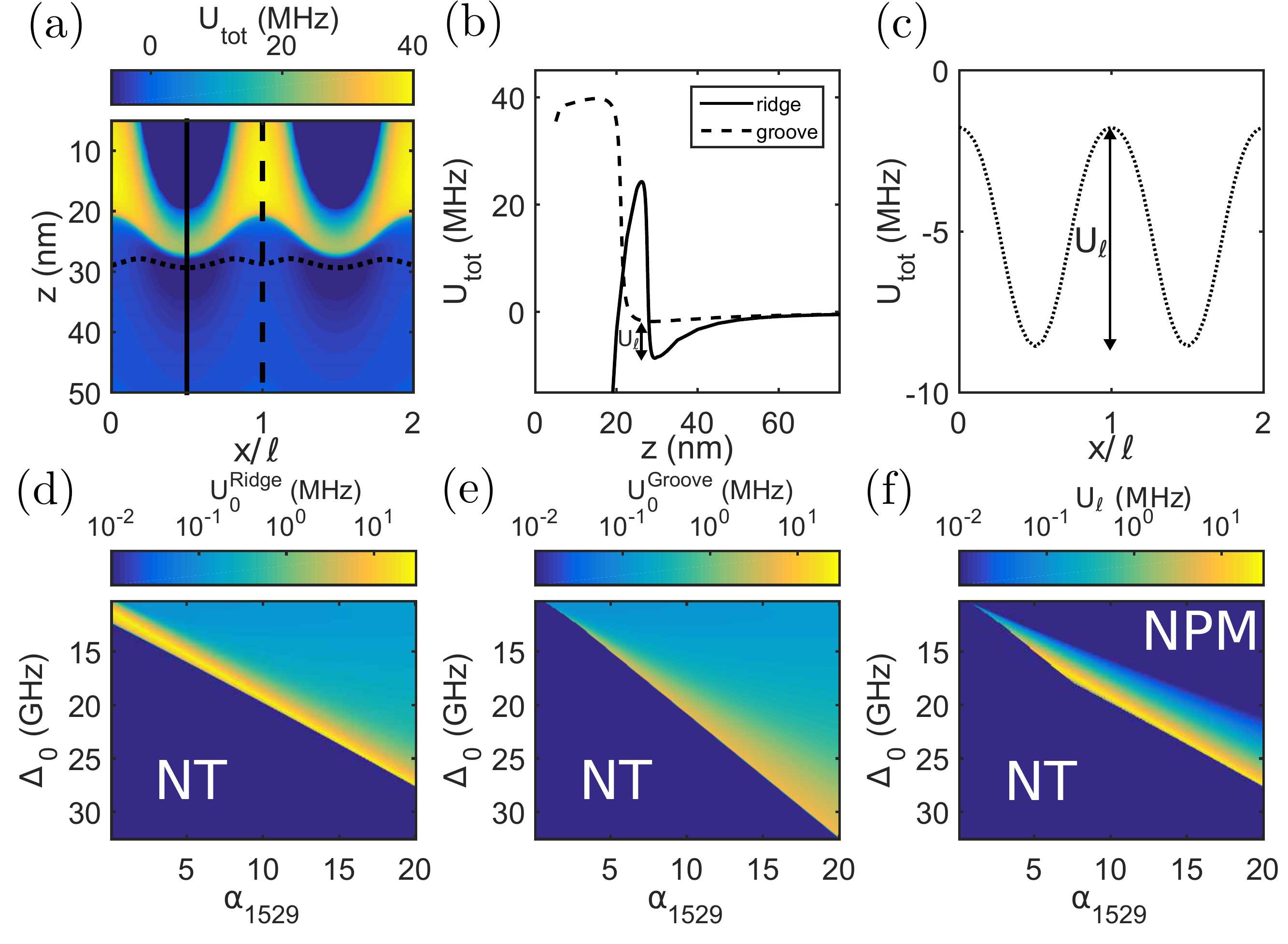}
\caption{\textbf{1D lattice geometry.} \textbf{(a)} 2D directionnal plot of the total potential. The different lines represent \textbf{(b)} the longitudinal ($Oz$) total potential above a ridge (full line) and a groove (dashed line) and \textbf{(c)} the transverse ($Ox$) total potential (dotted line). The atoms are trapped in a lattice at $z_t=29$nm  with a lattice depth $U_\ell=6.8$~MHz and a period $\ell=100$nm, for the following parameters : $P_{1529}^{Back}=500$mW, $\alpha_{1529}=7$, $\Delta_0/2\pi = 16.5$GHz, $P_{780}=0.1$mW ($\Omega_R/2\pi \sim 42$MHz). The trapping depths above \textbf{(d)} a ridge and \textbf{(e)} a groove create \textbf{(f)} a 1D lattice with a depth $U_\ell$ by tuning the $\alpha_{1529}$ parameter and the 780nm-laser detuning, with $\Omega_R/2\pi \sim 42$MHz. NT = Non Trapped. NPM = Non Periodically-Modulated.}
\label{fig6}
\end{figure}

The profile of the total trapping potential is plotted in Fig.\ref{fig6}\textbf{(a)} with longitudinal and transverse cross-sections in Fig.\ref{fig6}\textbf{(b)} and \textbf{(c)} respectively. Here, the parameters are $P_{1529}^{Back}=500$mW, $\alpha_{1529}=7$, $\Delta_0/2\pi = 16.5$GHz and $\Omega_R/2\pi \sim 42$MHz. In this configuration, the atoms are trapped in front of a ridge at $z_t\sim 29$~nm and the modulation of the ground-state CP potential creates lattice and trap depths of $\left\{U_{\ell},U_{0}\right\}=\left\{6.8,8.6\right\}$~MHz ($U_{\ell}\sim 118E_R$ with $E_R$ the lattice recoil energy) and a period of $100$~nm. The trapping frequencies are $\left\{\omega_x,\omega_z\right\}=2\pi\cdot\left\{6,32\right\}$~MHz. In Fig.\ref{fig6}\textbf{(d)}-\textbf{(f)}, the $\alpha_{1529}$ parameter is scanned to find the optimal intensity profiles that create a trap both above a ridge and a groove (Fig.\ref{fig6}\textbf{(d)},\textbf{(e)}).\footnote{We note that the existence of a lattice potential requires the formation of a trap in front of a ridge and the presence of a barrier in front of a groove.} The final lattice depth, which corresponds to the energy difference between the groove and ridge energy minima, is plotted in Fig.\ref{fig6}\textbf{(f)}.  As shown in Fig.\ref{fig6}\textbf{(d)}-\textbf{(f)}, for $\alpha_{1529}>7$ the lattice depth is little sensitive to an increase of the 1529nm-forward-laser power ($P_{1529}^{Front}$). To minimize the heating effect due to laser power, we choose $\alpha_{1529}=7$, which fixes the spatial profile of the light intensity while keeping its amplitude as a free parameter \textit{via} the power of the 1529nm-backward-laser beam ($P_{1529}^{Back}$).

\begin{figure}[h]
\centering
\includegraphics[scale=0.5]{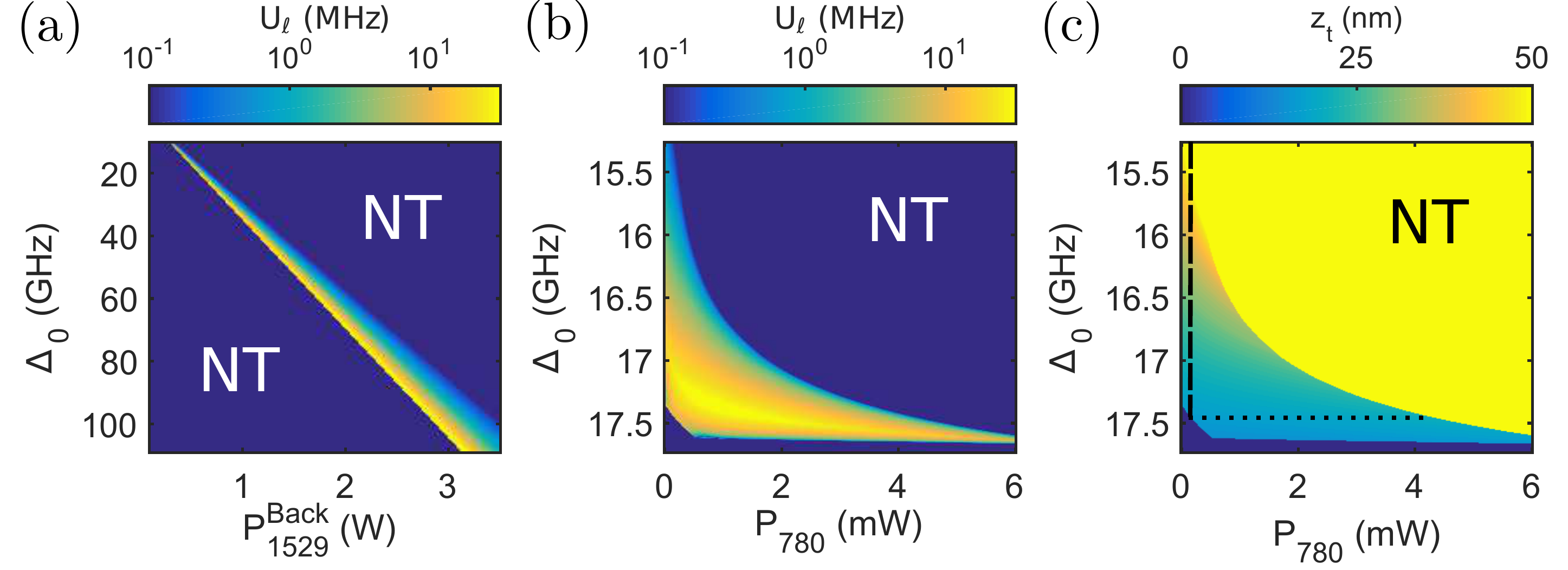}
\caption{ \textbf{Scanning the laser intensities.} Lattice depth $U_\ell$ \textit{versus} the 780nm-laser detuning $\Delta_0$ and \textbf{(a)} the 1529nm-backward-laser power $P_{1529}^{Back}$ and \textbf{(b)} the 780nm-laser radiation power $P_{780}$. \textbf{(c)} Scan of the trapping postion $z_t$ \textit{vs} the free parameters of the 780nm laser (detuning and power). Dashed line : $z_t$ varies from 17 to 60~nm by tuning $\Delta_0$ from 17.45 to 15.27~GHz at constant $P_{780}=0.2$~mW; Dotted line : $z_t$ varies from 17 to 22~nm by tuning $P_{780}$ from 0.2 to 4.2~mW at constant $\Delta_0=17.45$~GHz. NT = Non Trapped. Here we took : $\alpha_{1529}=7$ for all panels and $P_{1529}^{Back}=500$~mW for (b) and (c).}
\label{fig7}
\end{figure}

As demonstrated before, the $P_{1529}^{Back}$ parameter is key to increase the atom lifetime in the DDS trap (Fig.\ref{fig4}). In Fig.\ref{fig7}\textbf{(a)} we show the dependence of the lattice depth as a function of $P_{1529}^{Back}$. The 780nm-laser beam can also be used to change the trap parameters by controlling the coupling between the ground ($5S$) and excited ($5P$) states. As shown in Fig.\ref{fig7}\textbf{(b)}, by tuning both the frequency and the power of the 780nm laser, the one-dimensionnal lattice depth can be tuned from 0 to $\sim 35$ MHz (610 $E_R$). Although the contribution of these parameters to the lattice depth seems equivalent (symmetry), their influence on the trap-to-surface distance is very different (Fig.\ref{fig7}\textbf{(c)}). Indeed, a scan of $P_{780}$ at constant $\Delta_0$ has little effect on $z_t$ (dotted line) while a scan of $\Delta_0$ at constant $P_{780}$ shifts the trap from 17 to 60~nm (dashed line).

\section{\label{sec:level3}Discussion}

We have described and analyzed an experimental method to trap atoms in the near field of planar and nanostructured surfaces. The properties of the trap formed with this method can be tuned in lattice and trapping depths over relevant scales to simulate Hubbard Hamiltonians (0 - 600 $E_R$). The lattice parameters are experimentally adjusted by tuning the amplitude and frequency of laser beams, which are variables very well-controlled in laboratories. The surface geometry studied here led to a one-dimensional lattice. The idea can be straightforwardly extended to generate 2D subwavelength lattices at the cost of an increased numerical complexity for the computations of the optical response of the 2D structured surface. The algorithm for the 2D calculations based on Rigorous Coupled-Wave Analysis (RCWA) has already been developed \cite{moharam,Lalanne98} and is available on software such as Reticolo \cite{reticolo}.

In \cite{chang14}, a similar trapping method has been proposed in which the repulsive shape of the 6P$_{3/2}$ state of Cesium was induced by a Drude material with plasmon resonance frequency corresponding to the atomic transition and a Q-factor of $10^7$. This condition is extremely stringent on material engineering and a practical implementation seems hardly possible. Our  method is an interesting compromise where the excited-state properties are tuned by simple optical means which are little influenced by the material properties. Regardless of experimental implementation issues, we point out that \cite{chang14} benefits from a $z^{-3}$ divergence in potential close to the surface, which allows to access a diverging gradient of the excited state close to the surface. On the contrary, our method has an exponentially-decaying intensity profile and its slope is therefore upper-bounded by its value at the dielectric-vacuum interface. To trap atoms at distance below 15 nm, our method will require unrealistic optical power or close to transition frequency that will make it experimentally inappropriate.

In our method, the constraint for the choice of materials are guided by: a high refractive index in order to generate strong surface plasmon waves, very low absorption at 1529 nm so as to handle high optical powers and the existence of nanofabrication protocole to generate the desired nanostructures. In this work we have shown the predicted potentials for $Si$ and $SiO_2$ materials, which have negligible absorption at 1529 nm\footnote{We could not find any measured absorption at this wavelength in the literature where it is estimated to have zero absorption.} as well as a reasonably high refractive index ($n_{Si}=3.48$). In addition, the geometry of nanostructures that we have detailed in this work can be realized by Metal Assisted Etching (MAE). This technique allows for creating nanopillar array with dimensions down to a few tenth of nanometers in diameters and up to a micron in height \cite{kara16}. MAE can be extended to other types of material. Further improvement can be carried on the material choice to optimize the trapping geometry.

We have seen in section \ref{stratPart} that the atomic lifetime in the trap was of the order of 100 ms for the chosen range of parameters. This time was influenced by the exponentially-decaying optical profile from the dielectric-vacuum interface. In the 1D grating case, this exponentially-decaying profile was found to be absent in front of a groove and was compensated by the forward laser. The latter reduces the slope in front of a ridge that in turn limits the anti-damping time to 1 ms at best for our parameter range. Improvement in the lifetime in the lattice trap requires further investigation. In particular, the optimization of the geometry was carried by maximizing the intensity gradient in front of a ridge, which is crucial to increase the lattice depth. In future work, it would be interesting to consider as well the gradient of intensity in front of a groove and to look at different angles of incidence for the laser beams to improve the overall slope and consequently the lifetime of atoms in this subwavelength lattice. Another possibility to improve the lifetime would be to add some laser tunability to the long-range CP attraction force.  For example, this can be achieved by a red-detuned evanescent field \cite{Mildner18} that would help to increase the lattice depth at large distances to maintain long lifetimes.

\section{\label{sec:level4}Conclusion and perspectives}

In this work, we have presented the doubly-dressed state trapping method and applied it to trap atoms in the near field of various surfaces. We emphasized specifically on planar stratified structure and on 1D periodically-structured surface to generate periodic subwavelength potentials. For both cases, we have shown that the trap characteristics (trap-to-surface distance, depth and trapping frequencies) could be adjusted on a large range by simply changing the frequency and/or power of the two dressing lasers. In the case of the stratified medium, we simulated the different contributions to the trap loss and showed that trap lifetime of $\sim$100 ms could be expected. For the case of 1D nanostructures, we calculated the modulation of the Casimir-Polder potential and took advantage of this modulation to generate a subwavelength lattice. We showed that the lattice and trap depths could be adjusted by controlling the dressing-laser parameters. We anticipate that our demonstration of a 100 nm lattice spacing can be extended to smaller sizes as low as a few tenth of nanometers below which the optical power to form a trap will become irrelevant for applications. Our study was developed and adapted to $^{87}$Rb but it can be straightforwardly applied to most alcaline atoms for which the excited-state transition lies in the range 1-1.6 $\mu$m where the absorption of $Si$ is low. Other materials such as $GaP$ could also be exploited as it has a high refractive index and a very low absorption above 550 nm. In addition, materials such as $GaP$ could be interestingly used to include other optical functionalities on the surface such as optical waveguides for readout and addressing or surface guided modes for optically-mediated long-range interactions \cite{hood16,gonzalez15}. The reduction of lattice spacing is one key possibility to increase energies and to speed up dynamics in future cold-atom lattice experiments.

In the present work, the numerical tools that have been used are  well adapted to periodic structures but the use of Casimir-Polder potential structuration goes beyond this periodic frame. By nanoshaping the surface, we can access a large variety of lattice geometries. The possibilities include the design of pillar defects to simulate the role of impurities in solid-state physics, the introduction of a  controlled disorder or confined optical modes \cite{douglas15}. The Doubly-Dressed State method can be directly applied to tapered nanofibers, hollow-core fibers and slow light waveguides for which near-field trapping will strongly increase the atom-photon coupling strength.

%------------------------------------------------

\section*{Acknowledgements}

We thank Romain Pacanowski for numerical assistance. J.Z. acknowledges support from China Scholarship Council (CSC) and La Fondation Franco-Chinoise pour la Science et ses Applications (FFCSA). P.B. thanks the r\'egion Aquitaine : chaire of excellence.

\section*{Competing financial interests}

The authors declare no competing financial interests.

\end{document}